\documentclass[authoryear,preprint,review, 12pt]{elsarticle}
\usepackage{amssymb}
\usepackage{amsthm}
\usepackage{amsmath}
\usepackage{amsfonts}
\usepackage{amsbsy}
\usepackage{mathtools}
\usepackage{nicefrac}
\usepackage{bm}
\usepackage{url}  
\usepackage[utf8]{inputenc}
\usepackage[american]{babel}
\usepackage{mathtools}

\usepackage{color}
\definecolor{gray}{rgb}{0.5,0.5,0.5}

\journal{Statistics and Probability Letters}

\def\M1{{\mathcal M}_1}

\begin{document}

\begin{frontmatter}

%% Title, authors and addresses

%% use the tnoteref command within \title for footnotes;
%% use the tnotetext command for theassociated footnote;
%% use the fnref command within \author or \address for footnotes;
%% use the fntext command for theassociated footnote;
%% use the corref command within \author for corresponding author footnotes;
%% use the cortext command for theassociated footnote;
%% use the ead command for the email address,
%% and the form \ead[url] for the home page:
%% \title{Title\tnoteref{label1}}
%% \tnotetext[label1]{}
%% \author{Name\corref{cor1}\fnref{label2}}
%% \ead{email address}
%% \ead[url]{home page}
%% \fntext[label2]{}
%% \cortext[cor1]{}
%% \address{Address\fnref{label3}}
%% \fntext[label3]{}

\title{Bayesian Estimation of Kendall's $\tau$ Using a Latent Normal Approach}

%% use optional labels to link authors explicitly to addresses:
%% \author[label1,label2]{}
%% \address[label1]{}
%% \address[label2]{}

\author[label1]{Johnny van Doorn\corref{cor1}}
\address{University of Amsterdam}
\cortext[cor1]{Correspondence concerning this article may be addressed to Johnny van Doorn, University of Amsterdam, Department of Psychological Methods, Valckeniersstraat 59, 1018 XA Amsterdam, the Netherlands. E-mail may be sent to	JohnnyDoorn@gmail.com. The project was supported partly by grant 283876 from the European Research Council. 
The authors thank two anonymous reviewers for their helpful comments and suggestions. Centrum Wiskunde \& Informatica (CWI) is the national research institute for mathematics and computer science in the Netherlands.}
\author[label1,label2]{Alexander Ly}

\author[label1]{Maarten Marsman}

\author[label1]{Eric-Jan Wagenmakers}

\address[label1]{University of Amsterdam}
\address[label2]{Centrum Wiskunde \& Informatica}

\begin{abstract}
%\40 WORDS: #now 35
The rank-based association between two variables can be modeled by introducing a latent normal level to ordinal data. We demonstrate how this approach yields Bayesian inference for Kendall's $\tau$, improving on a recent Bayesian solution based on its asymptotic properties.

\end{abstract}


\end{frontmatter}

\section{A Bayesian Framework for Kendall's $\tau$}
\label{intro}
Kendall's $\tau$ is a popular rank-based correlation coefficient. Compared to Pearson's $\rho$, Kendall's $\tau$ is robust to outliers, invariant under monotonic transformations, and has an intuitive interpretation \citep{Kendall1990}. 
Let $X = (x_1,...,x_n)$ and $Y = (y_1,...,y_n)$ be two data vectors each containing ranked measurements of the same $n$ units. For instance, $x$ could be the rank ordered scores on a math exam and $y$ the rank ordered scores on a geography exam, for $n$ test-takers. A \textit{concordant} pair is then defined as a pair of subjects $(i,j)$ where subject $i$ has a higher score on $x$ and $y$ compared to subject $j$, whereas a \textit{discordant} pair is defined as one where $i$ scores higher on $y$, but $j$ scores higher on $x$, or the other way around.   
Kendall's $\tau$ is defined as the difference between the number of concordant and discordant pairs, expressed as proportion of the total number of pairs:
\begin{equation}
\tau_{} = \frac{\sum_{1\leq i<j \leq n}^{n}Q((x_i,y_i),(x_j,y_j))}{n(n-1)/2},
\label{tau_eq}
\end{equation}
where the denominator is the total number of pairs and $Q$ is the concordance indicator function, which is defined by:
\begin{equation}
Q((x_i,y_i)(x_j,y_j)) =
\begin{cases}
-1& \text{if } (x_i - x_j)(y_i - y_j) < 0 \\
+1& \text{if } (x_i - x_j)(y_i - y_j) > 0 \\
\end{cases}.
\end{equation}
The function returns $-1$ if a pair is discordant, and returns +1 if a pair is concordant.
However, due to the nonparametric nature of Kendall's $\tau$ and the lack of a likelihood function for the data, Bayesian inference is not trivial. 

An innovative method for overcoming this problem was proposed by \cite{johnson_bayes_2005}, and involves the modeling of the test statistic itself, rather than the data. This method has been applied to Kendall's $\tau$ by \cite{yuan_bayesian_2008}, and was recently developed by \cite*{vanDoorn2018}. The inferential framework that follows from this work uses the limiting normal distribution of the test statistic $T^*$ \citep{Hotelling1936,noether_theorem_1955}, where 
\begin{equation}
T^* = \tau \sqrt{ \frac{9n(n-1)}{4n+10}}.
\end{equation}
Under $\mathcal{H}_0$, this limiting normal distribution is the standard normal, whereas under $\mathcal{H}_1$, this distribution is specified with a non-centrality parameter $\Delta$ for the mean, and a sampling variance of 1. 

However, the method  ---henceforth the \textit{original asymptotic method}--- might fall short on two counts.
Firstly, the asymptotic assumptions only hold for sufficiently large $n$ (i.e., $n$ $\geq$ 20, see \citealp{vanDoorn2018}). 
Secondly, the variance of the sampling distribution of the test statistic depends on the population value of Kendall's $\tau$. For $\tau = 0$, the sampling variance equals 1, but as $\mid \tau\mid  \rightarrow 1$, the variance decreases to $0$ \citep{Kendall1990,Hotelling1936}.

In the current article, we will explore two corrections that aim to improve Bayesian inference for Kendall's $\tau$:
\begin{enumerate}
	\item Within the asymptotic framework, the observed value of Kendall's $\tau$ can be used to set its sampling variance. We label this the \textit{enhanced asymptotic method}.
	\item Within a Bayesian latent normal framework, a latent level correlation is obtained and transformed to Kendall's $\tau$. We label this the \textit{latent normal method}.
\end{enumerate}

\section{Correction Using The Sample $\tau$}
\label{analyticTau}
A first correction to consider is to use the sample value of Kendall's $\tau$, denoted $\tau_{obs}$, to estimate the sampling variance of $T^*$, denoted $\sigma^2_{T^*}$. A convenient expression for the upper bound of $\sigma^2_{T^*}$ in terms of $\tau$ is given in \cite{Kendall1990}:
\begin{equation}
\sigma^2_{T^*} \leq \frac{2.5n(1-\tau^2)}{2n+5}.
\end{equation}
Using $\tau_{obs}$ as an estimate of $\tau$ provides a somewhat crude approximation to the sampling distribution of $T^*$. However, compared to using $\sigma^2_{T^*} = 1$ as in the original asymptotic method, working with the upper bound will result in a more narrow posterior for cases where $\tau \neq 0$. However, the enhanced asymptotic method still suffers from the use of asymptotic assumptions about the sampling distribution and variance of the test statistic.

\section{Correction Using The Latent Normal Approach}
\label{PCC}
\subsection{Latent Normal Models}
\begin{figure}
	\centering
	\centerline{
		\includegraphics[trim=0cm 0cm 0cm 0cm, clip=true,scale=1]{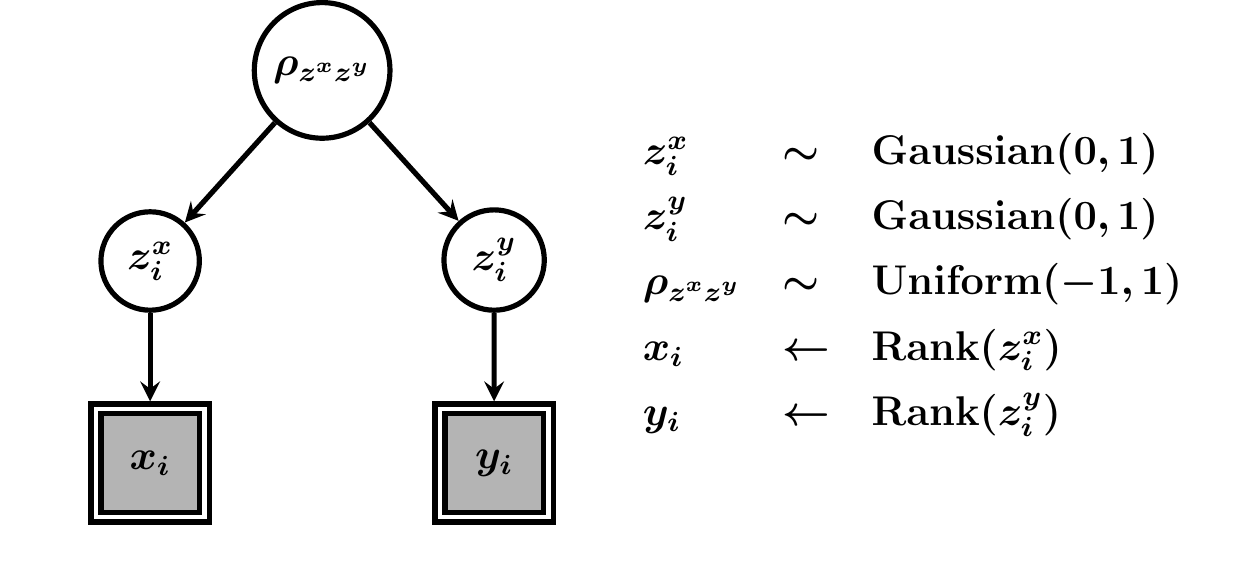}}
	\caption{A graphical model of the latent normal method. Here, $x$ and $y$ are observed rank data. The latent level is denoted with $z^x$ and $z^y$, and $\rho_{z^x z^y}$ represents the latent correlation.}
	\label{fig:graphmod}
\end{figure}

Several latent variable models quantify the association between two ordinal variables. These methods often introduce a latent bivariate normal distribution to the ordinal variables, where the association between variables is modeled through a latent correlation \citep{Pearson1900,Olssen1979,Pettitt1982,Albert1992,AlvoYu2014}. The observed rank data $(x,y)$ can then be seen as the ordinal manifestations of the continuous latent variables $(z^x,z^y)$, which have a bivariate normal distribution. Figure \ref{fig:graphmod} offers a graphical representation of such a model. Using this methodology, the nonparametric problem of ordinal analysis is transformed to a parametric data augmentation problem. 

%\begin{figure}
%	\centering
%	\centerline{
%		\includegraphics[trim=0cm 0cm 0cm 0cm, clip=true,scale=1]{Basic_TauPolychoric.pdf}}
%	\caption{A graphical model of the latent normal method. Here, $x$ and $y$ are observed rank data. The latent level is denoted with $z^x$ and $z^y$, and $\rho_{z^x z^y}$ represents the latent correlation.}
%	\label{fig:graphmod2}
%\end{figure}

%The latent normal method was first introduced by \cite{Pearson1900} as a means for modeling a $ 2 \times 2$ cross-classification table,
%and was further elaborated on by \cite{Pearson1922}, who coined the term \textit{polychoric correlation coefficient} (PCC) and accommodated analysis of $r \times s$ tables. A maximum likelihood approach for the PCC was developed by \cite{Olssen1979,Olssen1982}, and a Bayesian framework was introduced by \cite{Albert1992}.
%The PCC model is a special case of the ranked probit model where the number of categories is known. The method is also known as the Multivariate Normal Order Statistic model (MVNOS, \citealp{Yao1999, Yu2000}).

%In so-called rank likelihood models \citep{Pettitt1982, AlvoYu2014}, data augmentation is performed differently: the latent boundaries separating the rank categories are determined by the latent scores instead of estimated separately as in the MVNOS models.
%Both methods yield highly similar posterior distributions \citep{Hoff2007} and for simplicity we opted to use the rank likelihood methodology here.

\subsection{Posterior Distribution for the Latent Correlation}
\label{PostDist}

The joint posterior can be decomposed as follows:
\begin{equation}
P(z^x, z^y,\rho_{z^x, z^y} \mid  x,y) \propto P(x,y \mid z^x, z^y) \times P(z^x, z^y\mid \rho_{z^x, z^y}) \times  P(\rho_{z^x, z^y}).
\end{equation}
The second factor on the right-hand side is the bivariate normal distribution of the latent scores given the latent correlation:
\begin{equation}
\begin{pmatrix}z^x\\
z^y
\end{pmatrix} 
\sim  \mathcal{N}
\begin{bmatrix}
\begin{pmatrix}
0\\
0
\end{pmatrix}\!\!,&
\begin{pmatrix}
1 & \rho_{z^x, z^y}\\
\rho_{z^x, z^y} & 1\\
\end{pmatrix}
\end{bmatrix}.\\ ~\newline
\label{bivTruncNorm}
\end{equation}
The factor $P(x,y \mid z^x, z^y)$ consists of a set of indicator functions that map the observed ranks to latent scores, such that the ordinal information is preserved. For the value $z^x_i$, this means that its range is truncated by the lower and upper thresholds that are respectively defined as:
\begin{equation}
 a^x_i = \max\limits_{j: x_j < x_i}\left({z^x_j}\right) 
 \label{minthresh}
\end{equation}
\begin{equation}
 b^x_i = \min\limits_{j: x_j > x_i}\left({z^x_j}\right).
 \label{maxthresh}
\end{equation}

The third factor is the prior distribution on the latent correlation. In the remainder of this article, the prior is specified by a uniform distribution on $(-1,1)$ (but see \citealt*{Berger2008,LyEtAlsubm}).

The general Bayesian framework for estimating the latent correlation involves data augmentation through a Gibbs sampling algorithm \citep{Geman1984}, combined with a random walk Metropolis-Hastings sampling algorithm. At sampling time point $s$:
\begin{enumerate}
	\item For each value of $z^x_i$, sample from a truncated normal distribution, where the lower threshold is $a^x_i$ given in \eqref{minthresh} and the upper threshold is $b^x_i$ given in \eqref{maxthresh}:
	
	\[ (z_i^x \mid  z_{i'}^{x}, z_i^y, \rho_{z^x, z^y}) \sim N\left(z_i^y \rho_{z^x, z^y}, \sqrt{1-\rho^2_{z^x, z^y}}\right)\delta_{a^x_i,b^x_i} \]
	
	\item  For each value of $z^y_i$, the sampling procedure is analogous to step 1.
	\item Sample a new proposal for $\rho_{z^x, z^y}$, denoted $\rho^*_{z^x, z^y}$, from the asymptotic normal approximation to the sampling distribution of Fisher’s z-transform of $\rho$ \citep{Fischer1915}:
	\[ \tanh^{-1}(\rho^*_{z^x, z^y}) \sim N\left(\tanh^{-1}\left(\rho^{s-1}_{z^x, z^y}\right), \frac{1}{\sqrt{(n-3)}}\right) .\] 
	 
	The acceptance rate $\alpha$ is determined by the likelihood ratio of $(z^x,z^y \mid  \rho^*_{z^x, z^y})$ and $(z^x,z^y \mid  \rho^{s-1}_{z^x, z^y})$, where each likelihood is determined by the bivariate normal distribution in \eqref{bivTruncNorm}:
	\[ \alpha = \min\left(1,  \frac{P(z^x,z^y \mid  \rho^*_{z^x, z^y})}{P(z^x,z^y \mid \rho^{s-1}_{z^x, z^y})}  \right). \]
	
\end{enumerate}
Repeating the algorithm a sufficient number of times yields samples from the posterior distributions of  $z^x, z^y,$ and $\rho_{z^x, z^y}$.

\subsection{Relation to Kendall's $\tau$}
\label{Greiners}
% The next step in estimating Kendall's $\tau$ involves taking the posterior samples of $\rho_{z^x, z^y}$, and applying Greiner's relation 
With the posterior distribution for the latent $\rho_{z^x, z^y}$ in hand, the transition to the posterior distribution for Kendall's $\tau$ can be made using Greiner's relation
\citep{Greiner1909, Kruskal1958}. This relation, defined as
\begin{equation}
\label{Greiner}
\tau = G(\rho) = \frac{2}{\pi} \sin^{-1}(\rho)
\end{equation} 
enables the transformation of Pearson's $\rho$  to  Kendall's $\tau$ when the data follow a bivariate normal distribution.

Using Greiner's relation, the posterior distribution of Kendall's $\tau$ can be rewritten as follows:
\[ P(\tau \mid  x,y) \approx P(G(\rho) \mid  x, y ) = \int\int P(G(\rho) \mid  z^x,z^y) P(z^x,z^y \mid  x,y) dz^x dz^y .\] 
%\[ P(\tau \mid  x,y) \approx P(G(\rho) \mid  x, y ) = \left\mid \frac{dG(\rho)}{d\rho}\right\mid  \int\int P(G(\rho) \mid  z^x,z^y) P(z^x,z^y \mid  x,y) dz^x dz^y .\] 
Introducing the latent normal level to the observed variables enables the link between Pearson's $\rho$ and Kendall's $\tau$, and turns posterior inference for Kendall's $\tau$ into a parametric data augmentation problem that can be solved with the above MCMC-methods. Thus, Greiner's relation can be applied to the posterior samples of $\rho_{z^x, z^y}$ to yield posterior samples of $\tau_{}$.

Furthermore, the application of Greiner's relation in this manner implicitly alters the prior from a uniform distribution on the latent correlation to the following distribution on Kendall's $\tau$:
\begin{equation}
p(\tau) = \frac{\pi}{4} \cos \left( \frac{\pi \tau}{2} \right) \text{, for } \tau \in (-1\text{, }1).
\label{Phi_Tau}
\end{equation}

\section{Results: Simulation Study}
The performance of the original asymptotic method, the enhanced asymptotic method, and the latent normal method was assessed with a simulation study. \textcolor{black}{ For four values of $\tau$ (0, 0.2, 0.4, 0.7) and three values of $n$ (10, 20, 50), $10{,}000$ data sets were generated under four copula models: Clayton, Gumbel, Frank, and Gaussian \citep{Sklar1959,Nelsen2006,Genest2007,Colonius2016}. Using Sklar's theorem, copula models decompose a joint distribution into univariate marginal distributions and a dependence structure (i.e., the copula). The aforementioned copulas are governed by Kendall's $\tau$, so  the performance of each method can be assessed through a parameter recovery simulation study.  Furthermore, the univariate marginal distributions can be transformed to any other distribution using the cumulative distribution function and its inverse. Because these functions are monotonic, this does not affect the copula or ordinal information in the synthetic data and therefore vastly increases the scope of the simulation study. }
%Because copulas model model only the dependence structure and not the marginal distribution of the data, copulas can be used to cover a wide array of types of data. }

\begin{figure}[h!]
	\centering
	\centerline{
		\includegraphics[trim=0cm 0cm 0cm 1cm, clip=true,scale=0.33]{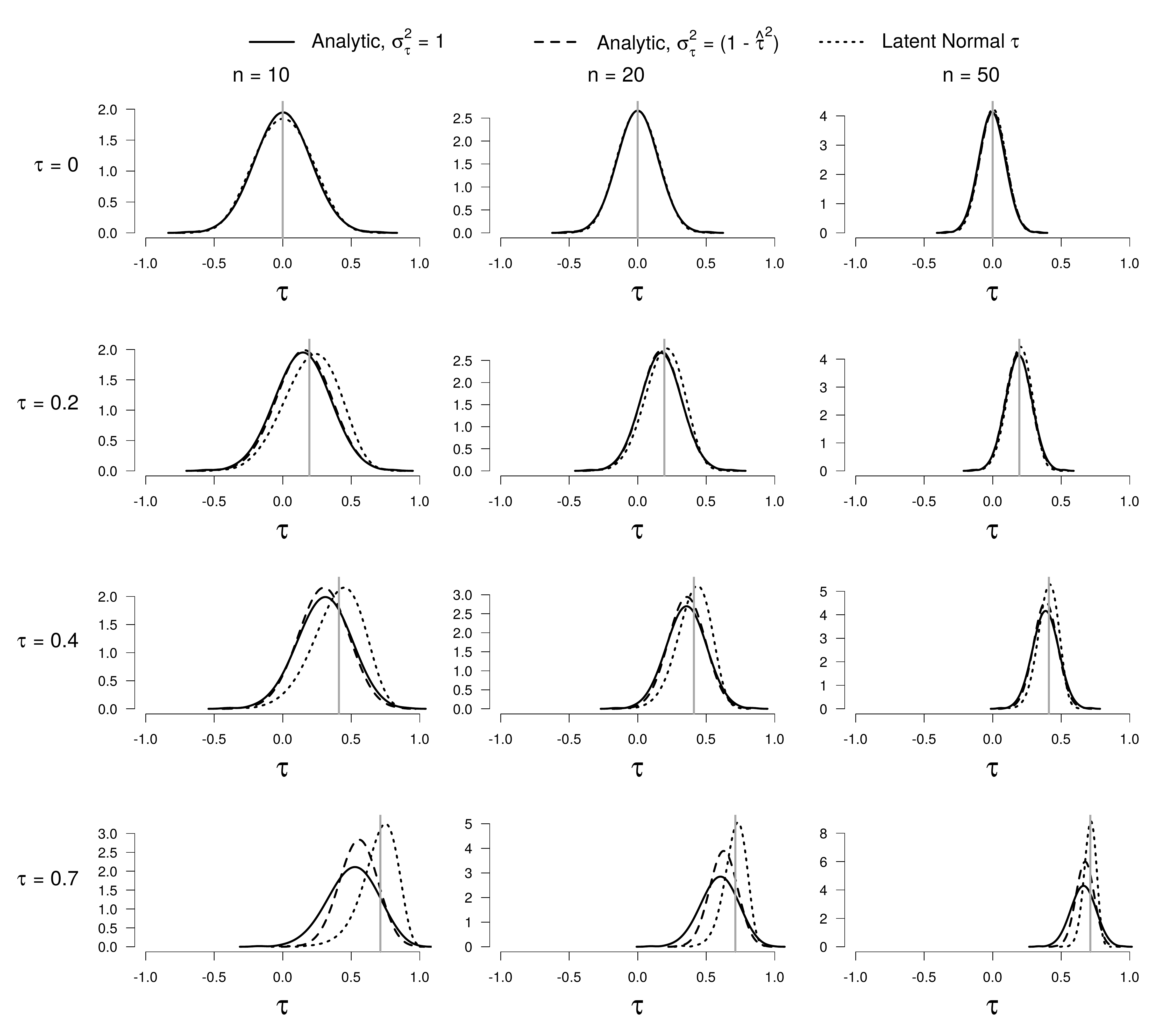}}
	\caption{To illustrate the performance of the three methods, quantile averaged posterior distributions for several values of $\tau$ and $n$ are shown. Each column corresponds to a value of $n$, and each row corresponds to a value of $\tau$. The quantile averaged posterior distributions were obtained with 10,000 synthetic datasets per combination of $n$ and $\tau$. The vertical gray line indicates the population value of $\tau$.}
	\label{fig:avQplot}
\end{figure}

For each data set, a posterior distribution was obtained using the three methods and the population value of $\tau$ was estimated using the posterior median. 
Per combination of $n$ and $\tau$, this resulted in $10,000$ posterior distributions. For an overall view of each method's performance, Figure \ref{fig:avQplot} shows the quantile averaged posterior distributions, along with a vertical line indicating the population value of $\tau$. \textcolor{black}{The data in Figure \ref{fig:avQplot} were generated using the Clayton copula; other copula models yielded highly similar results}. 
The quantile averaged posteriors indicate no difference between the inferential methods under $\mathcal{H}_0$, which corroborates the assumption of $\sigma^2_{T^*} = 1$ when $\tau = 0$. 
However, the difference in methods becomes pronounced in the scenario where $n = 10$ and $\tau = 0.7$. Both asymptotic approaches show a degree of underestimation, and yield a relatively broad posterior distribution. 
In the panels where $\tau \neq 0$, the misspecification of the sampling variance also becomes clear, as it is overestimated and results in a wider posterior distribution compared to the latent normal method. \textcolor{black}{ Although the assumption of latent normality is the price to pay for the Bayesian latent normal methodology, the simulation results indicate robustness of the method to various violations of this assumption.\footnote{\textcolor{black}{R-code, plots, and further details of the simulation study are available at \url{https://osf.io/u7jj9/}.}}}

\section{Discussion}
This article has outlined two methods of improving the Bayesian inferential framework in cases where $n$ is low and/or $\tau$ is high. Although an extension of the asymptotic framework performs somewhat better than the original asymptotic framework in \cite{vanDoorn2018}, both are outperformed by the latent normal approach. Under $\mathcal{H}_0$, the methods do not differ from each other, underscoring the validity of the general framework.

\textcolor{black}{The outlined methods are useful for both estimation and hypothesis testing. In the former case, the posterior distribution enables point estimation through the posterior median, or interval estimation through the credible interval. For hypothesis testing, the Savage-Dickey density ratio \citep{DickeyLientz1970,wagenmakers_bayesian_2010} can be used to obtain Bayes factors \citep{KassRaftery1995}. A concrete example is presented in the online appendix. 
Because the method uses only the ordinal information in the data, it retains the robust properties of Kendall's $ \tau$, such as invariance to monotone transformations, robustness to outliers or violations of normality, and ability to detect nonlinear monotone relations.  }

%% If you have bibdatabase file and want bibtex to generate the
%% bibitems, please use
%%

\section{Literature}
\bibliographystyle{elsarticle-harv}

\bibliography{/home/johnny/GoogleDrive/MyBibTex/myLibrary.bib}

%% The Appendices part is started with the command \appendix;
%% appendix sections are then done as normal sections
\appendix

\end{document}